\documentclass[aip,cha]{revtex4-1}
\usepackage{graphicx}
\usepackage{dcolumn}
\usepackage{bm}
\usepackage{amssymb}
\usepackage{pifont}
\usepackage{amsmath}
\usepackage{times}
\usepackage{natbib}

\usepackage{epstopdf}
\usepackage{latexsym}
\usepackage{keyval}
\usepackage{ifthen}
\usepackage{moreverb}
\usepackage{color}

\begin{document}
\title{Blinking chimeras in globally coupled rotators}
\author{Richard Janis Goldschmidt}
\affiliation{Department of Physics and Astronomy,
University of Potsdam, Potsdam 10623, Germany}
\affiliation{Institute of Pure and Applied Mathematics,
       University of Aberdeen, Aberdeen AB24 3FX, UK}
\author{Arkady Pikovsky}
\affiliation{Department of Physics and Astronomy,
University of Potsdam, Potsdam 10623, Germany}
\affiliation{Department of Control Theory,
Lobachevsky University Nizhny Novgorod,
Nizhny Novgorod 603022, Russia}
\author{Antonio Politi}
\affiliation{Institute of Pure and Applied Mathematics,
       University of Aberdeen, Aberdeen AB24 3FX, UK}

\date{\today}

\begin{abstract}
In globally coupled ensembles of identical oscillators so-called chimera states
    can be observed.
The chimera state is a symmetry-broken regime, where a subset of oscillators forms a
    cluster, a synchronized population, while the rest of the system remains a
    collection of non-synchronized, scattered units.
We describe here a blinking chimera regime in an ensemble of seven globally
    coupled rotators (Kuramoto oscillators with inertia).
It is characterized by a death-birth process, where a long-term stable cluster
    of four oscillators suddenly dissolves and is very quickly reborn with a new,
    reshuffled configuration.
We identify three different kinds of rare blinking events and give a quantitative
    characterization by applying stability analysis to the long-lived 
    chaotic state and to the short-lived regular regimes which arise when the cluster
    dissolves.
\end{abstract}

\maketitle

\begin{quotation}
Coupled oscillators can synchronize if the coupling is attractive, or
    desynchronize if the coupling is repulsive.
This basic effect is captured by the famous Kuramoto-Sakaguchi model of phase
    oscillators.
However, if inertia is included, i.e.\ the units are rotators,
    more complex regimes in between synchrony and asynchrony can be observed.
One such regime is a chimera pattern, where some rotators form a fully
    synchronous, perfect cluster, while the others are non-synchronized and
    all mutually different.
In this paper we report one such a chimera characterized by an interesting additional
    long-time dynamics.
We call it a \emph{blinking chimera} because 
every once in a while an event occurs where the cluster opens up and
quickly closes into a new reorganized composition. This event takes place on a time scale much shorter
    than that of the very long chaotic transient that is the chimera pattern --- the system \emph{blinks}.
We describe in detail how the exchange between the cluster and desynchronized
    units takes place.
	
\end{quotation}

\section{Introduction}
In the last years coupled oscillators proved to exhibit a very rich
variety of regimes, ranging from perfect synchronization to extremely
homogeneous asynchrony.
The most intriguing regimes are the intermediate ones, especially
when the oscillators spontaneously split into distinct groups/clusters.
Among them, chimera states are currently attracting a large interest.
They are characterised by the coexistence of synchronized and desynchronized groups
of identical oscillators which, in spite of their indistinguishability, do not all
behave in the same way.
The first chimera was
discovered by Kuramoto and Battogtokh~\cite{kuramoto_coexistence_2002}
in a one-dimensional medium of nonlocally coupled phase oscillators.
Since then, many setups have been found, where symmetry is broken, giving rise
to the simultaneous presence of synchronous and asynchronous subsets
(see the reviews~\cite{panaggio_chimera_2015} for both theoretical analyses
and the description of experimental setups).

Furthermore, examples of chimera regimes within rather small sets of
oscillators have been reported~\cite{maistrenko_smallest_2017}.
Recent studies have revealed that chimera states can be quite
complex. In particular, in breathing
chimeras~\cite{kemeth_classification_2016,bolotov_breather_2017,%
bolotov_simple_2018,suda_breathing_2018},
some oscillators join and leave the synchronous domain
because of oscillations of the governing order parameter.

In the following sections we report on a different steady nonstationary
chimera. We describe a situation where a well-defined chimera persists for
a very long time and is suddenly destroyed; shortly afterwards, a new reshuffled
and long-lived chimera reforms.
Since the reshuffling events are rather short compared to the long chimera stages, we call this state
\emph{blinking chimera}. 
Noteworthy, in the dynamics of networks, the notion
of blinking systems is well established~\cite{bl,bl2}. There, switchings in the coupling and/or
network topology are imposed according to a pre-defined external protocol ---
for example periodic or random blinking.
In our case, the blinking events are not pre-described, but appear spontaneously 
due to the dynamical rules.

The paper is organized as follows: we introduce the model in Sec.~\ref{sec:model}, and
describe the phenomenology of blinking events in Sec.~\ref{sec:ph}. Next, we
develop a quantitative characterization of blinking in Sec.~\ref{sec:qbe}. We conclude
with a discussion in Sec.~\ref{sec:con}.

\section{The basic model}\label{sec:model}
The Kuramoto-Sakaguchi model (KS model from now) of globally coupled phase oscillators is a widely used
system to study synchronization phenomena,
see reviews~\cite{acebron_kuramoto_2005,gupta_kuramoto_2014,pikovsky_dynamics_2015}. For
identical units, the KS model is exactly
solvable~\cite{watanabe_integrability_1993,watanabe_constants_1994,pikovsky_dynamics_2015},
yielding either complete synchronization (if the coupling is attractive)
or an asynchronous state with vanishing order parameter (if the coupling is
repulsive). The exact solution shows that chimera states, characterized by the
coexistence of fully synchronous (identical) oscillators with asynchronous
units, cannot arise.

Integrability breaks up if the original model is perturbed.
One popular extension of the KS model consists in including
the effects of inertia on the oscillating units, i.e. in replacing phase oscillators
with rotators~\cite{tanaka_first_1997,hong_inertia_1999,dorfler_synchronization_2013,%
filatrella_analysis_2008, grzybowski_synchronization_2016,ha_large_time_2014,%
barre_bifurcations_2016, belykh_bistability_2016,Olmi-15}:
\begin{equation}
\alpha \ddot{\phi}_i + \dot{\phi}_i = \frac{1}{N} \sum^N_{j=1}
\sin\left(\phi_j-\phi_i-\beta\right)\;.
\label{eq:ki}
\end{equation}
Here $\phi_i$ is the instantaneous phase of the \(i\)-th rotator,
$N$ is their number,
\(\beta\) is a constant phase shift (sometimes called \emph{frustration}
in the literature), and, finally,
$\alpha$ is the (dimensionless) mass of the rotators.
A constant torque acts on all rotators and the equations
are written in the reference frame rotating with the corresponding
constant frequency (thus the torque does not enter).

At this point it is instructive to discuss an essential difference between oscillators and rotators.
For example, pendula can perform both oscillations and rotations. When they are used in pendulum clocks
and in metronoms~\cite{Martens_etal-13,sr}, they operate as self-sustained oscillators. For these
oscillations, a phase can be introduced, which is different from the angle variable $\phi$ in \eqref{eq:ki}
and obeys a first-order (in time) equation. Thus, for coupled oscillators (and for all setups of coupled
metronoms) the usual first-order Kuramoto model is appropriate; the model ``with inertia''
cannot be used for oscillators, but for rotators only.

Systems of coupled rotators have been widely discussed in the literature. 
In Refs.~\onlinecite{tanaka_first_1997,PhysRevE.90.042905}, diversity of torques has been
shown to result in a hysteretic transition to synchrony. Effects of noise and diversity
have been treated analytically and semi-analytically in
Refs.~\onlinecite{PhysRevLett.81.2229,PhysRevE.62.3437,PhysRevE.89.022123}. One of the popular 
applications of the rotator model of type \eqref{eq:ki} are power grids~\cite{dorfler_synchronization_2013,%
filatrella_analysis_2008,PhysRevLett.109.064101, grzybowski_synchronization_2016,%
PhysRevE.90.042905,PhysRevLett.109.064101,nc}.
In these applications one does not consider a mean field coupling like in Eq.~\eqref{eq:ki}, but a network
of different producers and consumers of electrical power, with different values of torque and 
different connectivities. Another much studied setup is that of
symmetric deterministic models where chimera states emerge as a result of symmetry breaking.
This is the case of a one-dimensional medium with nonlocal coupling studied in Ref.~\onlinecite{PhysRevE.91.022907}
and of two symmetric subpopulations with different couplings (within and between them)
considered in Ref.~\onlinecite{PhysRevE.92.030901}.
Here, we consider a setup that is even ``more" symmetric, since all pair-wise interactions are equal to one 
another. The phenomena we thus observe are entirely due to the prescribed dynamics and cannot be 
attributed to diversity among the oscillators, network topology, or noise.

For $\alpha \to 0$ the system~\eqref{eq:ki} reduces to the standard
KS system, describing the behavior
of (identical) phase oscillators. Therefore,
for small $\alpha$-values, the dynamics is expected to closely reproduce
that of the KS model.
In the limit $\alpha \to \infty$, the system converges to the
Hamiltonian mean field model~\cite{gupta_kuramoto_2014}: a paradigmatic
model for the study of long-range interactions in the presence of
a conservative dynamics.

We expect interesting and potentially new phenomena to arise
in the region where attraction and repulsion nearly balance each other.  This
is indeed the parameter region where standard chimera states are observed.
More specifically, we have selected $\beta = 0.53\cdot\pi$, which corresponds
to a weakly repulsive coupling in the KS model, while in the KS model with
inertia neither the fully synchronous nor the splay states are stable.
Additionally, in order to investigate the role of inertia, we have selected a
finite and relatively large mass $\alpha=10$.
As for the number of oscillators, we assume $N=7$: it is the smallest 
system size for which blinking chimeras
have been observed.  For $N<7$ we have observed only either
simple clustered or fully disordered states.

\section{Phenomenology}
\label{sec:ph}
In this section we qualitatively describe chimera states and their blinking;
a quantitative characterisation and a more thorough analysis is
postponed to the next section.

The equations of motion~\eqref{eq:ki} have been simulated by
implementing a standard 4th
order Runge-Kutta integrator with a timestep of \(\Delta t= 0.01\).
Phases \(\phi_i\) and frequencies \(\dot{\phi}_i\) are initialized
by drawing them from random distributions,
\(\phi_i \in \mathcal{U}(0,2 \pi)\), and \(\dot{\phi}_i \in \mathcal{N}(0,1)\).
Moreover, we have introduced numerical inhomogeneities on the mass \(\alpha\)
of the
order \(\Delta \alpha \simeq O\left(10^{-14}\right)\) to prevent
the oscillators \emph{clumping together} due to finite floating point precision.
We classify the current configurations by identifying clusters of oscillators
in almost identical states.
Two oscillators indexed by \(i\) and \(j\) are said to belong to the same cluster
when their distance $d(i,j) < 10^{-10}$, where
\begin{equation}\label{eq:dist}
d(i,j) = \sqrt{\delta(\phi_i-\phi_j)^2+(\dot \phi_i - \dot \phi_j)^2}
\end{equation}
and $\delta = \min(|\phi_i-\phi_j|,2\pi-|\phi_i-\phi_j|)$ to take into account that $\phi_i$ is equivalent
to $\phi\pm 2\pi$.

\subsection{Instability of the splay and the fully synchronous states}
Before describing the formation of the chimera state in detail, we comment on the
instability of the completely synchronous and the fully-asynchronous (splay)
solutions. The completely synchronous state is given by $\phi_i = \phi_j$ and
$\dot\phi_i = \dot\phi_j, \: \forall i, j$, where all oscillators collapse into a cluster.
We study the stability of this cluster via the transversal Lyapunov exponent
(which tells us whether a virtual pair of oscillators would be repelled or attracted by
the cluster, see Sec.~\ref{sec:qbe} and Eq.~\ref{eq:dcl} below for a
discussion). The resulting second order equation for the
transversal perturbation $\alpha\ddot\delta+\dot\delta+\cos\beta \delta$ 
can be solved analytically.  The exponents are
$\lambda_{1,2}=(-1\pm\sqrt{1-4\alpha\cos\beta})/(2\alpha)$, yielding 
for the parameters under consideration
$\lambda_1 \approx 0.047$ and $\lambda_2 \approx -0.147$. Hence, the fully synchronous cluster is unstable with transversal Lyapunov exponent
$\lambda \approx 0.047$.

The splay state is characterized by the frequencies $\dot\phi_i = 0, \forall
i$, and an equidistribution of phases on the unit circle, $\phi_i = i\cdot2\pi/N, \: i\in\lbrace 0, N-1\rbrace$,
$N=7$ being the number of oscillators in the system. It is difficult to analyse the instability of this steady
state analytically, but numerical exploration is straightforward: one easily observes
that a small (of order $10^{-12}$) perturbation grows exponentially with the
exponent $\lambda \approx 0.1$. Thus, both the fully synchronous cluster and the splay state are unstable.


\subsection{Formation of a chimera state}
\label{sec:fch}
\begin{figure}[!htb]
\includegraphics[width=0.55\columnwidth]{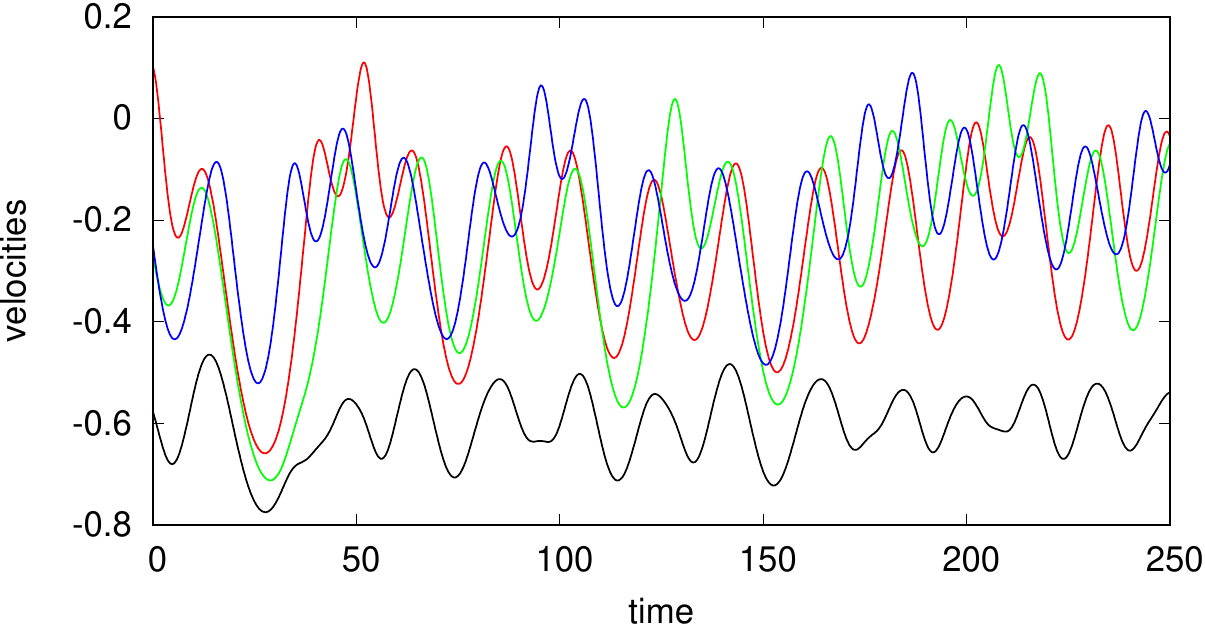}\hfill
\includegraphics[width=0.42\columnwidth]{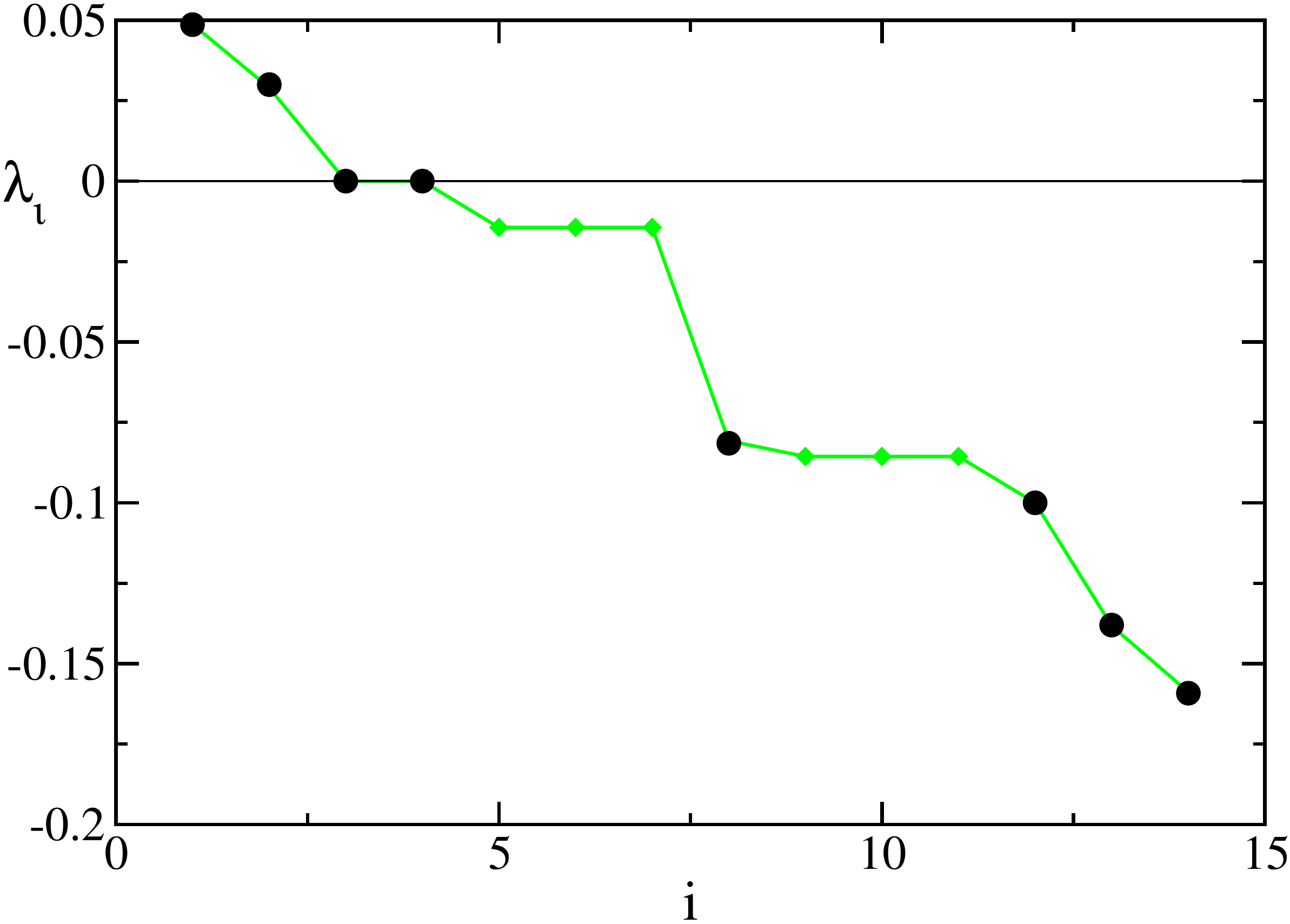}
\caption{Left panel: time series of rotator velocities in the chaotic chimera
state. Red, blue and green curves: isolated (not belonging to a cluster) units, black:
cluster of 4 units. Right panel: the full Lyapunov spectrum in the chaotic regime (green
symbols, partially overlapped with black ones).
It has two positive Lyapunov exponents
(thus this regime can be characterised as hyperchaos),
two zero LEs (due to two invariances - with respect to shift of time and with
respect to shift of all phases), with all other exponents being negative. One can
see a degeneracy due to the presence of a cluster: there are two groups of three equal
LEs ($i=5,6,7$ and $i=9,10,11$),
which correspond to the transversal directions of the cluster, see discussion
of the transversal LEs below. Other exponents (black circles) coincide with those
of the reduced system~\eqref{eq:kir}, where the cluster configuration
\texttt{4-1-1-1} is fixed so that the system is 8-dimensional and has 8 LEs.}
\label{fig:chch}
\end{figure}

The free evolution from random initial conditions leads
to the formation of a chimera state, where four oscillators clump together
to form a cluster, while the three other oscillators remain isolated from one another (we indeed
refer to them as to \emph{isolated units}). We denote this chimera state as \texttt{4-1-1-1}.
It appears that this state is a global attractor, as in our 
numerical simulations we never observed other configurations formed from random initial conditions.
Chimera states of this type have been observed
in globally coupled identical phase oscillators with delay and
in globally coupled Stuart-Landau
oscillators~\cite{sethia_chimera_2014,schmidt_coexistence_2014,yeldesbay_chimeralike_2014}.
The average time for the formation of a chimera is $\approx 1.8\cdot10^{3}$.
The corresponding dynamics is chaotic, as qualitatively visible in the left panel of
Fig.~\ref{fig:chch}, where we plot the time series of the rotator velocities $\dot\phi_i$.
 Here all units are chaotic: those belonging to the cluster and the isolated ones.
On a more quantitative level, the corresponding Lyapunov spectrum is plotted in the right panel.
It is composed of 14 exponents, two of which are indeed larger than zero, two vanish
(due to invariance under time translation and under a homogeneous shift of the phases), while
all the others are negative.

In the spectrum, we also notice two triples of identical negative exponents.
As confirmed below (see Sec.~\ref{sec:qbe}), they account for the transversal stability
of the 4-cluster.
The remaining eight Lyapunov exponents (see the filled black circles in Fig.~\ref{fig:chch})
contribute to the dimension of the underlying attractor. By virtue of the Kaplan-Yorke formula,
$D_{KY}\approx 4.8$ represents an upper bound to the information dimension.

No other configurations have been observed in the system~\eqref{eq:ki}
for the same parameter values --- neither fully synchronous states, nor chimera-type configurations
with 2, 3, 5, or 6 elements in the cluster.

\subsection{Blinking of chimera}

\begin{figure}[!htb]
        \includegraphics[width=0.5\columnwidth]{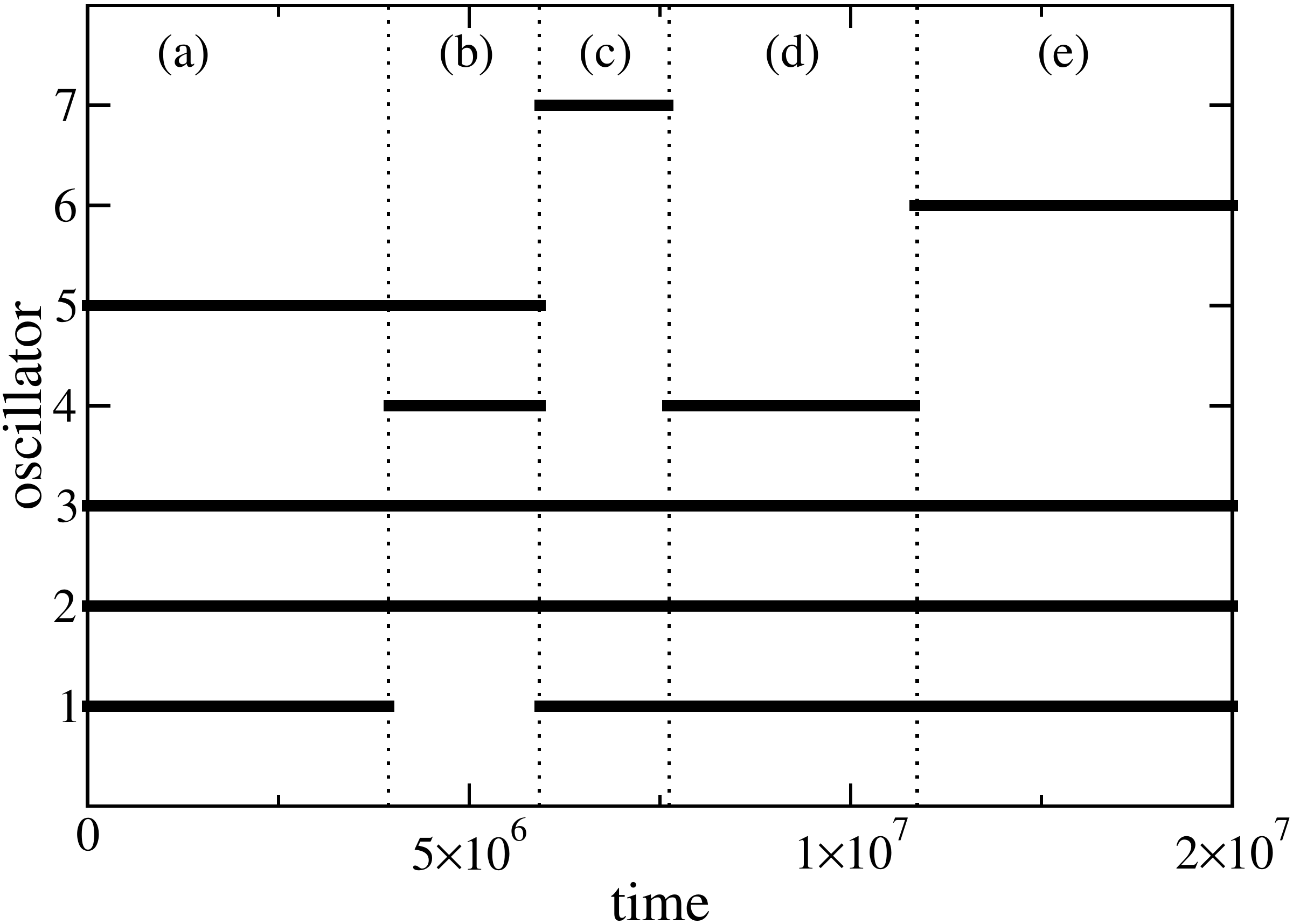}
        \caption{Pattern of the oscillators belonging to the
        four-oscillator cluster (marked by black squares which are seen as bold lines) versus time.
        One can clearly see five epochs with different cluster compositions. In epoch (a), oscillators 1, 2, 3, and 5 belong
        to the four-cluster. In epoch (b), oscillators 2, 3, 4, and 5. In (c), oscillators 1, 2, 3, and 7. In (d), oscillators 1, 2, 3, and 4.
        And finally in epoch (e), oscillators 1, 2, 3, and 6 belong to the cluster.}
\label{fig:pattern}
\end{figure}

\begin{figure}[!htb]
\centering
\includegraphics[width=0.7\columnwidth]{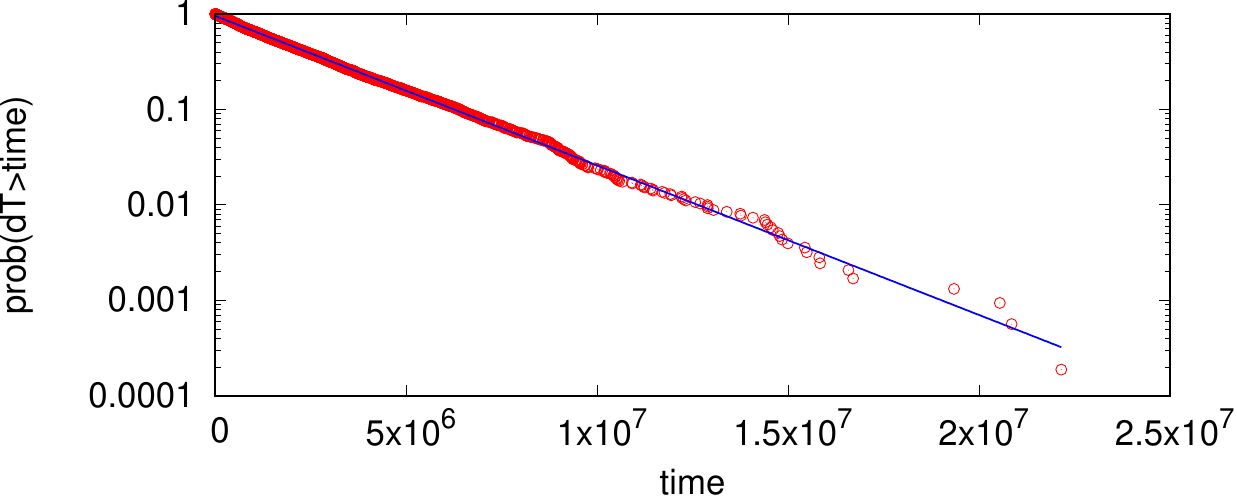}
\caption{Distribution of time intervals between different reshuffling events}
\label{fig:ti}
\end{figure}

The \texttt{4-1-1-1} regime described above is observed on a relatively long time scale, but it
is not the asymptotic one. Over very long time scales, one observes a picture
like in Fig.~\ref{fig:pattern}: rare events lead to a reshuffiling of
the cluster composition, with some oscillators leaving the cluster and others joining. These reshufflings
are observed continuously and they are apparently independent random
events following a Poisson process with a rate $\approx 3.6\cdot 10^{-7}$ (corresponding
to a mean time $\approx 2.8 \cdot 10^6$ between the events).
This follows from an exponential distribution of inter-event time intervals,
presented in Fig.~\ref{fig:ti}.

The observed pattern of reshuffling events shows that the chimera
state is not stationary, but blinking. Below we provide further illustrations
of this blinking, and characterise it more quantitatively.

\subsection{Destruction and re-formation of a 4-1-1-1 chimera state}
\label{sec:dr}

\begin{figure}[!htb]
    \centering
   \includegraphics[width=0.5\columnwidth]{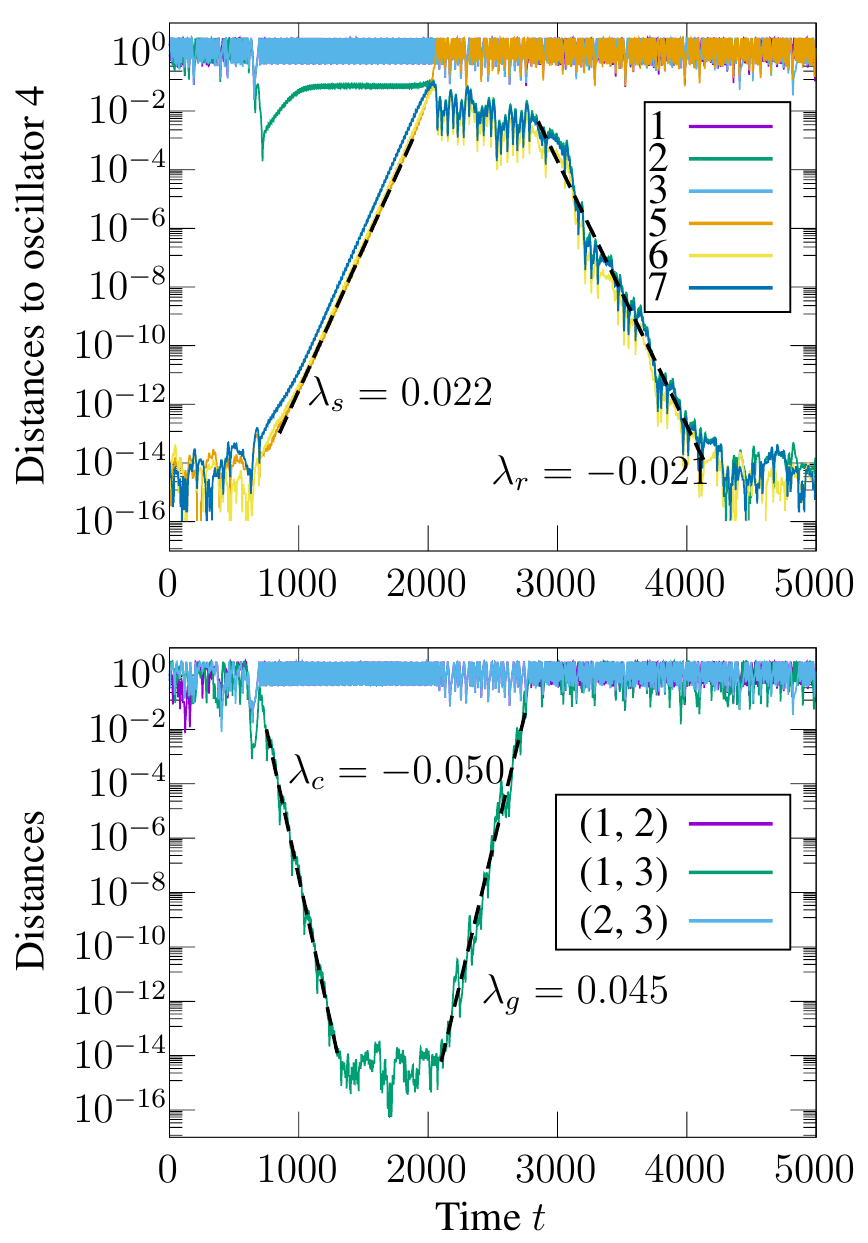}
    \caption{%
    Reshuffling of a \texttt{4-1-1-1} chimera state: scenario A.
    Top panel: distances of all oscillators from oscillator 4.
    Isolated oscillators at the beginning are 1, 2, 3,
    distances between them are depicted in the bottom panel. Around $t\approx 700$,
    a regular regime emerges from the chaotic \texttt{4-1-1-1} state,
    with one isolated oscillator (here unit 2). Simultaneously,
    the 4-cluster begins to
    disintegrate (with a rate $\lambda_s$) and two of the initially
    isolated units come close and form a 2-cluster state (here units 1,3).
    (bottom panel,
    rate $\lambda_c$). Disintegration ends at $t\approx 2000$
     in a chaotic state, where the units 2, 4, 6, 7 form a not so perfect 4-cluster
     (mutual distances are $\approx 10^{-2}$), while the units 1, 3 form a
     2-cluster (see bottom panel),
     and unit 5 is isolated. The 2-cluster begins to disintegrate with a rate
     $\lambda_g$; this disintegration ends around $t\approx 2800$.
    From this moment on, the new 4-cluster becomes more stable,
    the mutual distances reduce with the rate $\lambda_r$.
    At the end of the event, at $t\approx 4000$,
    a reshuffled configuration \texttt{4-1-1-1} appears.
    }
    \label{fig:rs1}
\end{figure}

\begin{figure}[!htb]
    \centering
  \includegraphics[width=0.5\columnwidth]{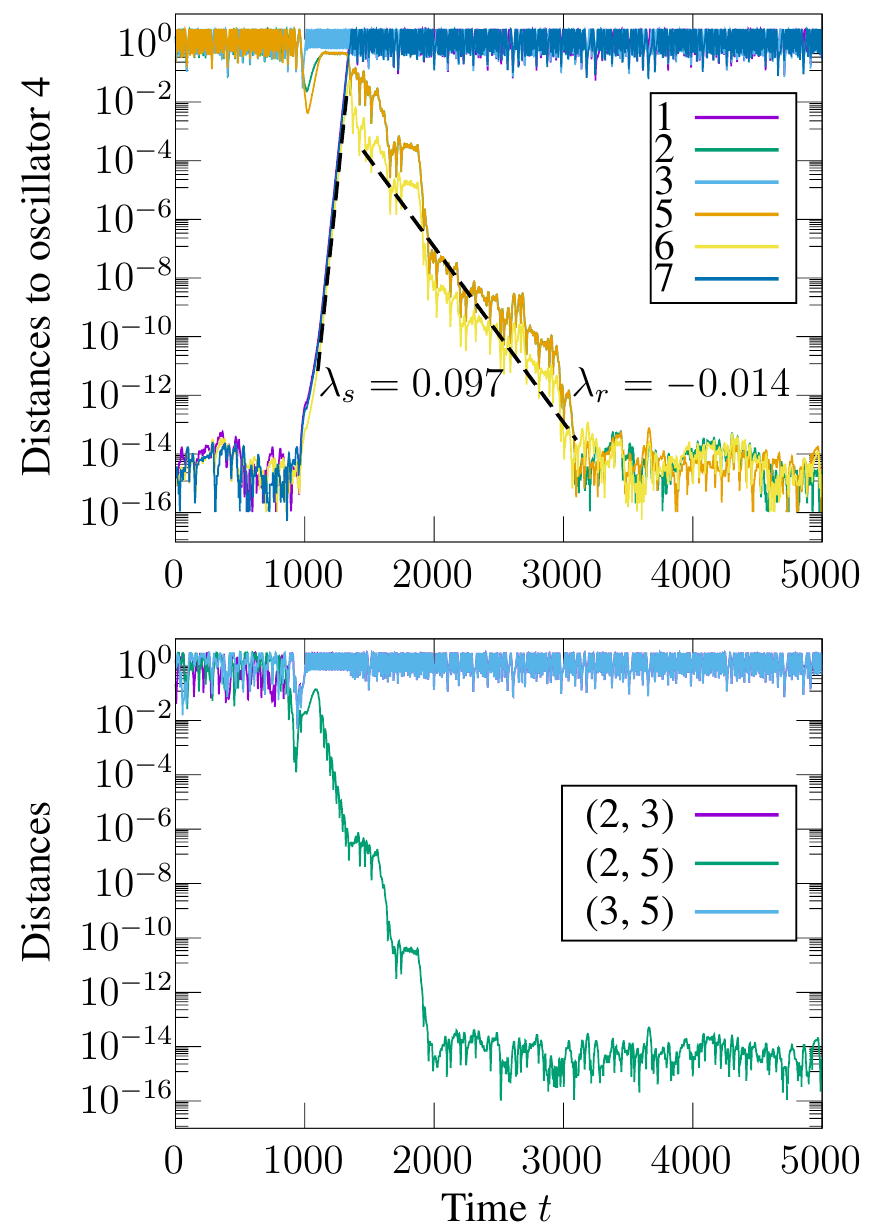}
    \caption{%
        Reshuffling of a \texttt{4-1-1-1} chimera state: scenario B.
    Top panel: distances of all oscillators from oscillator 4. Isolated
    oscillators at the beginning are 2, 3, 5; distances between them are depicted
    in the bottom panel. The process is qualitatively similar to
    that of Fig.~\ref{fig:rs1}, but here in the regular stage  around
    $t\approx 1000$ two isolated oscillators (2 and 5)
    are both close to each other and to the
    still existing cluster; they eventually join the new 4-cluster.
    }
    \label{fig:rs2}
\end{figure}

\begin{figure}[!htb]
    \centering
  \includegraphics[width=0.5\columnwidth]{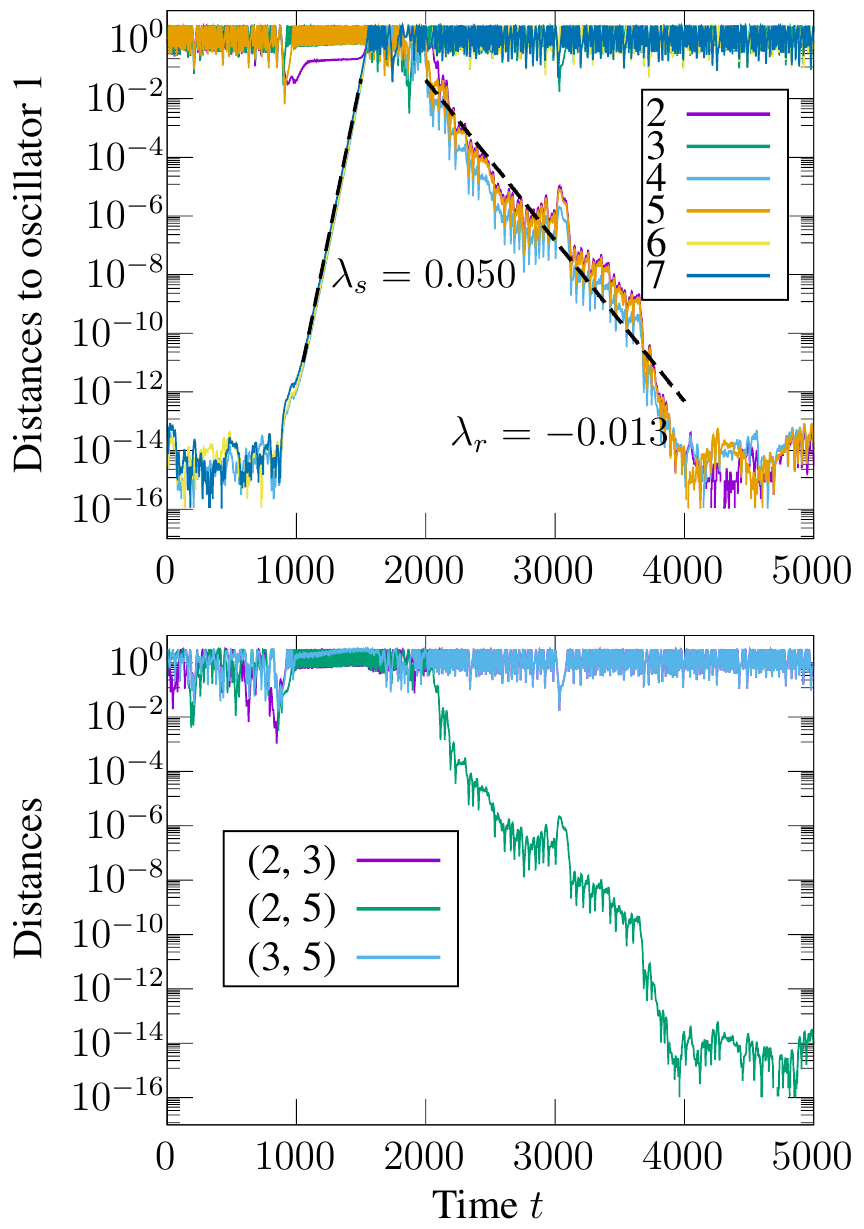}
    \caption{%
        Reshuffling of a \texttt{4-1-1-1} chimera state: scenario C.
    Top panel: distances of all oscillators from oscillator 1. Isolated
    oscillators at the beginning are 2,3,5; distances between them are depicted
    in the bottom panel. At the end units 1,2,4,5 belong to the 4-cluster
    and units 3,6,7 are isolated.
    }
    \label{fig:rs3}
\end{figure}

Here we describe in detail, mostly qualitatively, what happens
during the blinking (reshuffling) events. In fact, we have found that
three different reshuffling scenarios can happen; they are presented in
Figs.~\ref{fig:rs1}, \ref{fig:rs2}, and \ref{fig:rs3}.
We refer to these cases as to A, B, and C. We
first describe scenario A with reference to Fig.~\ref{fig:rs1}; it will
then be straightforward to explain also the other two scenarios.

In the two panels of Fig.~\ref{fig:rs1} we show distances between oscillators,
defined according to Eq.~\eqref{eq:dist}, as a function of time
(time offset is chosen arbitrarily at some instant around 1000 units prior the start
of reshuffling). In this event, the initial 4-cluster configuration contains the units 4, 5, 6,
and 7, while the reshuffled 4-cluster contains the units 2, 4, 6, and 7. Accordingly, in the top panel
we depict distances from unit 4, which belongs to the cluster both prior and after reshuffling
(in the same way, we always choose the reference unit as belonging to the cluster
prior and after the event in top panels of Figs.~\ref{fig:rs2},\ref{fig:rs3}).
In the bottom panel we show distances between all the three pairs of units not belonging to the cluster
prior to reshuffling.

In the top panel of Fig.~\ref{fig:rs1}, one can appreciate the presence of a chaotic \texttt{4-1-1-1} state
for $t\lesssim 800$. It is followed by a regular regime during which the isolated unit 2
comes rather close to the 4-cluster. In this regime, the 4-cluster is unstable and starts
dissolving. Meanwhile, the units 1 and 3 come close to each other. The dissolution of the 4-cluster,
accompanied by the appearance of a 2-cluster emerging from the isolated units,
continues until $t\approx 2000$. Afterwards, the 2-cluster dissolves and the dynamics
again become chaotic.
Around $t\approx 3000$, no clusters are observed.
Four units are close to each other, although they do not fulfill
our criterion for the definition of a cluster. However, they start
approaching each other and a new 4-cluster eventually forms
in the same way as observed when starting from random initial conditions
(cf. Section~\ref{sec:fch} above). Typically, the unit which was already close to the
4-cluster around $t\approx 1000$, joins the novel cluster, ``exchanging'' with one unit
that leaves the cluster.
However, in some cases the temporary 2-cluster does not dissolve but enters
the novel 4-cluster, exchanging with two units therein.

The blinking event shown in Fig.~\ref{fig:rs2} is quite similar to that of
Fig.~\ref{fig:rs1}, with the following differences: (i) now, during the regular
state arising at the beginning of reshuffling (around $t\approx 1000$),
not one, but two isolated units \emph{orbit} close to the 4-cluster (here units 2 and 5);
(ii) the disintegration of the 4-cluster is much faster than in Fig.~\ref{fig:rs1};
(iii) the two isolated units that were close to the 4-cluster join the new cluster,
so there is always a $2\leftrightarrow 2$ exchange.

The blinking event in Fig.~\ref{fig:rs3} is different from those in
Figs.~\ref{fig:rs1}, \ref{fig:rs2}: here there is no formation of
a temporary 2-cluster. The break-up of the 4-cluster (at $1000\lesssim t\lesssim 1600$)
is faster than in case Fig.~\ref{fig:rs1}, but slower than in case Fig.~\ref{fig:rs2}.
At the end of this process, all the units are separated, and a new
4-cluster begins to form,
with a $2\leftrightarrow 2$ exchange. Sometimes we observed that all 3 previously
isolated units
join the new cluster, so that also a $3\leftrightarrow 3$ exchange is possible.
On the other hand, since the formation of a new cluster is a statistical process,
it can happen that the final configuration is formed of the same units as the initial one.

\section{Quantifying blinking events}
\label{sec:qbe}

The above description of the reshuffling process suggests the existence of
regular (nonchaotic) temporary stages. To resolve them, we proceed by performing simulations
with four units \emph{glued} together to enforce the presence of a 4-cluster at all times.
Such a state (and, more generally, any clustered state) can be modelled also with
\textit{reduced} equations, the identical elements of a cluster being described
by just one set of variables $(\varphi,\dot\varphi)$. Generally, if $N$ rotators build
$K$ clusters of sizes $n_1,n_2,\ldots,n_K$, with $\sum_k n_k =N$,
then the equations \eqref{eq:ki} can be rewritten as a smaller reduced system of $K$ equations
\begin{equation}
\alpha \ddot{\varphi}_k + \dot{\varphi}_k = \text{Im}
\left(Z\mathrm{e}^{-i\varphi_k-i\beta}\right),\qquad Z=\sum_{k=1}^K \frac{n_k}{N}\mathrm{e}^{i\varphi_k}
\;.
\label{eq:kir}
\end{equation}
In our case of a \texttt{4-1-1-1} chimera, we have $K=4$ with $n_1=4,n_2=n_3=n_4=1$.
The reduced system \eqref{eq:kir} has the same dynamics as the original system
\eqref{eq:ki} so long as the 4-cluster persists.

In fact, simulations of the \texttt{4-1-1-1} chimera state with Eqs.~\eqref{eq:kir} reveal that this
regime is nothing but a very long pseudo-stationary chaotic transient~\cite{lai_transient_2011}.
The dynamics, eventually, collapses onto one of the following four attractors

\begin{description}
    \item[A] a quasiperiodic \texttt{4-2-1} state, characterized by a 2-cluster, while the isolated oscillator
    orbits close to the 4-cluster;
    \item[B] a quasiperiodic \texttt{4-2-1} state such that the 2-cluster orbits close to the 4-cluster;
    \item[C] a periodic \texttt{4-1-1-1} state, where three units remain asymptotically isolated;
    \item[D] a periodic \texttt{4-3} state.
\end{description}

Upon performing 1000 simulations of the reduced model, starting from random initial conditions,
we found that roughly 54\%, 31\%, 13\%, and 2\% converge to attractor A, C, B, and D, respectively.
We have used the notations A, B, C, because these regular states correspond to the scenarios
A, B, and C discussed in section~\ref{sec:dr}.
As for the probability to observe the various scenarios in the original blinking dynamics,
they are approximately equal to the above reported rates, the major difference being that
the state D has been never observed in the original system~\eqref{eq:ki}, presumably because
the 4-cluster is destroyed prior to the formation of the 3-cluster.

The reason why the attractors of the reduced model are not seen as such in the simulations of
the global system is that the 4-cluster is transversally unstable in all of the above scenarios.
Transversal Lyapunov exponents can be determined by perturbing only the variables of the cluster
$(\varphi_k,\dot\varphi_k)$, while leaving the mean field $Z$ unchanged
(cf. the general discussion of transversal Lyapunov exponents in
Ref.~\onlinecite{pikovsky_lyapunov_2016}). So the linear equation for
a deviation $(\delta_k,\dot\delta_k)$ from cluster $k$ reads
\begin{equation}
\alpha\ddot\delta_k+\dot\delta_k=-\delta_k\text{Re}[Ze^{-i\beta-i\varphi_k}] \; .
\label{eq:dcl}
\end{equation}
In practice, the above equation, solved together with Eq.~\eqref{eq:kir},
tells us whether a virtual pair of oscillators leaving cluster $k$ would be
either attracted or repelled by the cluster.
As Eq.~\eqref{eq:dcl} is two-dimensional, it yields two transversal Lyapunov
exponents; we are interested in the maximal one that can be
computed in the usual way by virtue of
the Benettin algorithm~\cite{pikovsky_lyapunov_2016}, i.e. by regularly renormalizing the
vector $(\delta_k,\dot\delta_k)$ and averaging the logarithm of the norm.

The implementation of this approach during the pseudo-stationary transient evolution of the \texttt{4-1-1-1}
configuration yields two values which coincide with the two triples visible in the general Lyapunov
spectrum plotted in Fig.~\ref{fig:chch} and confirm that the 4-cluster is stable on average.
In the case of the above mentioned four attractors, we instead find that the largest transversal
Lyapunov exponent is: $\lambda_t^A \approx 0.0227$ [for case A]; $\lambda_t^B \approx 0.1$ [B];
$\lambda_t^C \approx 0.05$ [C]; $\lambda_t^D \approx 0.17$ [D], thus confirming the instability
of the 4-cluster.

Regular attractors do not only qualitatively correspond to the initial states
of the blinking events described in Section~\ref{sec:dr}, but give also a quantitative
description of the disintegration of the 4-cluster: the growth rates
of the inter-oscillator
distances in Figs.~\ref{fig:rs1}, \ref{fig:rs2}, \ref{fig:rs3} correspond to the values
of the transversal Lyapunov exponents for cases A-C.

Additional insight into cases A and B can be obtained from the computation of
the transversal Lyapunov exponents of the 2-cluster in the corresponding \texttt{4-2-1} configuration.
In both cases its value is $\approx -0.05$.  This quantity
describes the rate $\lambda_c$ with which 2-cluster is formed, cf.
Figs.~\ref{fig:rs1} and \ref{fig:rs2}.

The formation of a new 4-cluster from a non-clustered chaotic regime is a
statistical event. However, its late stage, where the cluster is basically formed
and the units are progressively approaching each other, can be again compared with
the transversal Lyapunov exponents. In this context, the stable transversal
exponent $\lambda_t\approx -0.014$ of the 4-cluster in the chaotic regime (Fig.~\ref{fig:chch})
is relevant, as it gives approximately the convergence rate $\lambda_r$ in
Figs.~\ref{fig:rs1}, \ref{fig:rs2}, \ref{fig:rs3}.

Finally, the rate of disintegration of the two-cluster $\lambda_g$ (bottom
panel of Fig.~\ref{fig:rs1}) can be explained as follows: this regime corresponds to
a temporary chaotic state in the fixed configuration \texttt{4-2-1}. Calculations of
the transversal Lyapunov exponent of the 2-cluster here are not reliable,
as this regime quite soon ends in one of the A, B, D states. However,
if one starts the configuration \texttt{4-2-1} from random initial conditions,
in many runs one observes that during the initial stages the transversal Lyapunov exponent
of cluster 4 fluctuates around zero, while the transversal LE of
cluster 2 is $\approx 0.048$. This value corresponds to the rate $\lambda_g$
of disintegration of the 2-cluster in  Fig.~\ref{fig:rs1}.

Summarizing, we explained the origin of blinking events via an interplay of two properties of the system: structural,
i.e.\ the composition of clusters, and dynamical, i.e.\ the complexity of the dynamics and the resulting stability characteristics of
clusters. During long epochs a chaotic regime with stable \texttt{4-1-1-1}
clustering is observed. However, after a long but finite time, chaos is succeeded by a regular regime, and this
triggers a blinking event: first, the big cluster becomes unstable and dissolves, and then from a chaotic
disordered state a new chimera state with a reshuffled composition emerges.

\section{Discussion and conclusions}
\label{sec:con}

In this paper we reported on a novel state of blinking chimera in a small system of seven identical rotators 
(phase oscillators with inertia).
The asymptotically stationary regime consists of a sequence of long epochs each characterized by
a (temporary) chaotic chimera state with four oscillators synchronized into a single cluster,
and three isolated ones.
Such regimes are separated by relatively short reshuffling events when
the composition of the synchronous cluster is reconfigured.
These rare events, which we call \emph{blinking events}, appear to be distributed according to a Poisson
process with a very small, but finite rate. There are three types of such events (see the above described
scenarios A, B, and C); all of them are characterized by a regular
(either periodic or quasiperiodic --- depending on the scenario) dynamics.

Altogether, the chaotic chimera state is not an attractor, but rather a transient
chaotic state, eventually ending in a temporary regular dynamics. 
The emergence of long chaotic transients is a well known phenomenon in nonlinear dynamics~\cite{lai_transient_2011}:
they typically arise because of ``holes'' in phase space, where the 
trajectory suddenly jumps out of the pseudo-attractor.
Identifying the specific conditions for these events to occur is not an easy task: we leave
it to future investigations.
What makes the regime discussed in this paper different from standard chaotic transients is that
once the temporary chaotic state is over, another equivalent such regime emerges.
In fact, the three types of exit events all lead to unstable attractors.
In practice, during the blinking event, the cluster is transversally unstable and it thereby
starts disintegrating, leading to a short-lived non-chimera stage.
A new chimera configuration of the type \texttt{4-1-1-1} {\it finally} forms, due to the
transversal stability of this regime.

In order to clarify the quantitative properties of these processes, we explored the dynamics
of restricted systems with fixed cluster compositions. This allowed us to calculate, via time averaging,
the transversal Lyapunov exponents governing the stability of the clusters, without
destroying the clusters themselves. The fixed configuration \texttt{4-1-1-1} allowed us
to determine the basic unstable transversal Lyapunov exponents governing disintegration
of the main 4-cluster. In some cases, an intermediate 2-cluster is formed, with this
formation being governed by the stable Lyapunov exponent of the 2-cluster
in the fixed \texttt{4-2-1} configuration. Finally, the rate of formation of the
new 4-cluster from the unclustered chaotic state is governed by the stable
transversal Lyapunov exponent of the 4-cluster in the chaotic transient state of the \texttt{4-1-1-1}
configuration.

From the general viewpoint of topology of the dynamics in the
phase space of the system,
the blinking chimera can be described as follows: there is an invariant
manifold where the states of four oscillators coincide, while three differ.
This 8-dimensional
manifold is attractive for a set of large measures in the original 14-dimensional
phase space, but is not a global attractor. On this invariant set
a chaotic transient (chaotic saddle) sets in, characterized by a very long
lifetime: this regime corresponds to a chaotic chimera.
Eventually, generic trajectories leave the chaotic saddle and approach
one of the sets (A, B, C) all characterized by a regular dynamics.
On these regular sets the 8-dimensional manifold is transversally unstable, so that trajectories
leave it (the chaotic chimera is destroyed), but then enter again the domain
of attraction of the chaotic 8-dimensional saddle, and a new,
reshuffled chimera is established. Noteworthy, intermittent chaotic chimeras
have been reported for a two-population setup~\cite{PhysRevE.92.030901}. However, no reshuffling,
and thus also no blinking, was observed therein.

Preliminary simulations suggest that this phenomenon is not peculiar of the parameters selected
in this paper. However, in our simulations we never observed 
blinking chimera for less than seven units. 
Therefore, in this short communication we restricted ourselves to a description
of the minimal blinking chimera and do not discuss other possible dynamical states of this system.

\begin{acknowledgments}
We thank M. Rosenblum and Yu. Maistrenko for useful discussions.
This
work has been funded by the EU’s Horizon 2020 research
and innovation programme under the Marie 
Sklodowska Curie Grant Agreement No. 642563. A.Pik. acknowledges support
of the Russian Sceince Foundation (Grant 17-12-01534).
\end{acknowledgments}

%

\end{document}